\documentstyle[12pt,aasms4]{article}

\received{1995 July 28}
\accepted{January 3 1996}

\begin{document}

\title{LOOKING FOR THE S-Z EFFECT \\
TOWARDS DISTANT ROSAT CLUSTERS OF GALAXIES \altaffilmark{1}}

\author{P. Andreani\altaffilmark{2,7}}
\affil{Dip. di Astronomia, Padova, Italy}
\affil{European Southern Observatory, Garching, Germany}

\author{L. Pizzo\altaffilmark{3}}
\affil{Istituto TESRE, CNR, Bologna, Italy}

\author{G. Dall'Oglio\altaffilmark{4}}
\affil{Dip. di Fisica, III Univ. Roma, Italy}

\author{N. Whyborn\altaffilmark{5}}
\affil{S.R.O.N. Groningen, The Netherlands}
\author{H. B\"ohringer\altaffilmark{6}}
\affil{Max-Planck Institut f\" ur extraterrestrische Physik, Garching, Germany}
\author{P. Shaver\altaffilmark{7}}
\affil{European Southern Observatory, Garching, Germany}
\author{R. Lemke, A. Ot\`arola, L.-\AA.~ Nyman\altaffilmark{8}}
\affil{European Southern Observatory, La Silla, Chile}
\author{R. Booth\altaffilmark{9}}
\affil{ONSALA Space Observatory, Onsala, Sweden}

\altaffiltext{1}{Based on observations collected with the ESO-Swedish
SEST 15m telescope (La Silla, Chile).}
\altaffiltext{2}{Dipartimento di Astronomia, Vicolo Osservatorio 5, I-35122, 
Padova, Italy. e-mail: andreani@astrpd.pd.astro.it}
\altaffiltext{3}{Istituto TESRE, C.N.R., via P. Gobetti, Bologna, 
Italy. e-mail: pizzo@botes1.bo.astro.it}
\altaffiltext{4}{Dip. di Fisica, III Univ. Roma, P.le Aldo Moro
2, I-00185 Roma, Italy. e-mail: dalloglio@roma1.infn.it}
\altaffiltext{5}{S.R.O.N. Groningen, P.O. Box 800, Groningen, The Netherlands.
e-mail: nick@sron.rug.nl}
\altaffiltext{6}{Max-Planck Institut f\" ur extraterrestrische Physik,
Garching, Germany.}
\altaffiltext{7}{European Southern Observatory, Garching, Germany.
e-mail : pshaver@eso.org, pandrean@eso.org}
\altaffiltext{8}{European Southern Observatory, La Silla, Chile.
e-mail : rlemke@eso.org, aotarola@eso.org, lnyman@eso.org}
\altaffiltext{9}{ONSALA Space Observatory, Onsala, Sweden.}

\begin{abstract} 

We report on observations of the Sunyaev-Zeldovich effect towards
X-ray ROSAT clusters taken with a double channel (1.2 and
2 {\it mm}) photometer installed at the focus
of the 15m SEST antenna in Chile.
This paper describes the first results obtained for the high-z clusters
S1077, A2744 and S295.
Marginal detections were found for A2744 and at 1 {\it mm} for S1077.
We discuss these data in terms of
contamination of sources along the line of sight and give a constraint
on the amplitude of the kinematic effect.

\end{abstract}


\section{Introduction}
The Sunyaev-Zeldovich (SZ) effect is a characteristic signature on the Cosmic
Microwave Background (CMB) radiation spectrum originated from (inverse)
Compton
scattering of the photons by hot electrons in clusters of galaxies. There
has been considerable interest in its detection because of its potential
as a diagnostic tool for observational cosmology ({\it physical process
in the earlier universe; determination of H$_0$ and q$_0$ and peculiar
 velocities of clusters}) and for the study of the intracluster medium (see the
original papers by \cite{SZ69}, \cite{SZ72}, \cite{SZ81}).
 Its detection has been the
goal of many observational
searches (e.g.  \cite{B90}),
 mostly in the Rayleigh - Jeans (R-J) part of the spectrum,
where the scattering leads to an intensity decrement. However, interpretation
of measurements in the R-J side is plagued both by the possible 
radio emission from sources within the clusters and by the smallness of the
effect. After many attempts to detect this effect (see e.g., the review paper
by Birkinshaw 1990) radio observations show the expected
decrement at centimeter wavelengths
towards A2218, A665, 0016+16, A773 and Coma (\cite{B91}, \cite{KL91},
\cite{J93}, \cite{GR93}, \cite{HLR}) 
 and at 2.2{\it mm} towards A2163 (\cite{WIL94}).

\noindent
Measurements near the planckian peak and on the Wien side are in
principle  more
attractive since: (a) the intensity enhancement relative
to the planckian value is larger than the magnitude of the R-J decrement;
(b) the simultaneous detection of the enhancement (positive) and decrement
(negative) on the same cluster provides an unambiguous signature of
their presence;
(c) sources in the cluster are expected to give a negligible contribution at
high frequency; (d) because of the large bandwidth the sensitivity
of bolometer systems is excellent.

\noindent
We have built a photometer with two bolometers centered at 1.2
and 2 {\it mm} to feed the O.A.S.I. (Osservatorio Antartico
Submillimetrico Infrarosso) telescope installed at the Italian base in
Antarctica (\cite{DAL92}).
This photometer was adapted to the focus of the S.E.S.T.
and its performances were tested during an observing run
in September 1994.
Most of the allocated time was spent in installing
the photometer and testing the equipment (\cite{PIZ95a})
but some useful observations
were gathered towards S1077, A2744 and S295.

\section{The Instrument}

The instrument was described in detail elsewhere (\cite{PIZ95a}),
here only the main characteristics are summarized.
The photometer uses two Si-bolometers cooled to 0.3 K by means of
a single stage $^3$He refrigerator  (\cite{PIZ95b}).
Radiation coming from the telescope is focussed by a PTFE lens
and enters the cryostat through a
polyethylene vacuum window and two cooled filters.
A dichroic mirror
(beam-splitter) at 4.2 K
divides the incoming radiation between two f/4.3 Winston Cones cooled
at 0.3 K located orthogonal each other.
The wavelength ranges are defined by two interference filters centered at
1.2 and 2 {\it mm} cooled at 4.2K with bandwidth 350 and 560 $\mu m$
respectively.
The 2{\it mm} band includes the peak brightness of
the decrement in the
S-Z thermal effect, while the 1.2{\it mm} bandwidth is a compromise between
the maximum value of the enhancement in the S-Z and the atmospheric
transmission.

\noindent
Since our goal is the simultaneous observation of the same source region,
comparable beam sizes at the two different frequencies were achieved
using Winston cones with the same entrance sizes for both channels,
selected to have roughly
a diffraction limited configuration for the 2{\it mm} channel.
Beam shape and dimension were measured using
raster scans on Saturn, Jupiter and Uranus (details on the 
procedure used and on the map reconstruction are in Haslam (1974),
Sievers et al. (1991), Lemke (1992)).
The planet maps show a good alignment (within 2$^{\prime \prime}$)
of the two beams and
their symmetry around the optical axis. Their Half Power Beam Widths turn
out to be 44$^{\prime \prime}$ (1.2 {\it mm}
channel) and 46$^{\prime \prime}$ (2 {\it mm} channel).

\noindent
Since clusters of galaxies are extended sources, in order to have the
reference beam as far as possible from the peak region of the X-ray emission,
the chop throw was set to the maximum possible amplitude,
i.e. that corresponding to a beam separation on the sky of
135$^{\prime \prime}$.
The coupling between the antenna and the photometer was checked by
comparing the output signals from known sources at the
1.2{\it mm} channel of our photometer with
those at the 1.2{\it mm} MPI bolometer available at SEST
(\cite{KRE90}),
both having a filter very similar in transmission and shape.
Taking into account the correction factors due to the different
fields of view, equal signals within the errorbars have been found.

\noindent
The detector noise of both channels measured at the focus was about
50nV/$\sqrt{Hz}$. Responsivities were inferred with
Venus, Saturn, Uranus, Mars and Jupiter
taking the planet brightness temperatures from Ulich (1981 and 1984):
the {\it effective responsivities}
at the focus turned out to be 1.8$\mu$V/K and 0.7$\mu$V/K at 1.2 and
2 {\it mm} respectively.
This means that the measured values for the N.E.T.s
were 28 and 70 mK/$\sqrt{Hz}$, i.e. 14 and 35 mK in 1s
 integration time. 10 \% is the average uncertainty on these
figures.
Converting these values in thermodynamic temperatures gives for the
measured sensitivities ${\Delta T \over T} = 0.014$
and ${\Delta T \over T} = 0.02$ at 1.2 and 2 {\it mm} respectively
(1s integration time).

\noindent
Atmospheric transmission was monitored by frequent skydips, i.e. telescope
scans at constant azimuth and different elevation are carried out
to compare the atmospheric emission with that of an absorber
moved into the path of the reference beam.

\noindent
The position of the sources was found by first pointing
a nearby radio-loud
quasar with strong millimetric fluxes and then the X-ray
coordinates. With this procedure we reduce the pointing errors to less
than 2$^{\prime \prime}$.

\section{The observing strategy}

Targets were selected for their high X-ray luminosities in the ROSAT
band (0.1$\div$2.4 keV). They are extended
X-ray sources identified as optical high-redshift ($z \sim 0.3$)
clusters.  Names, redshifts and ROSAT luminosities are listed in table
1. If we assume a core radius of 250$\div$400 kpc in standard
cosmologies their apparent size is 45$\div$70$^{\prime \prime}$, a
chop throw of
135$^{\prime \prime}$ means that at the reference beam position the ratio
between the electron density, $n_e (\theta=135^{\prime \prime }$),
to its central value, $n_e(\theta=0)$, is $\sim 0.28\div0.42$, i.e.
we lose 30$\div$40
\% of the signal because of the limited chop throw.
The effective values of $y$, taking account of the SZ effect in the
off-center positions, can be roughly estimated only for A2744
for which a deep PSPC exposure is already available\footnote{ROSAT HRI
observations are already scheduled for S1077 and S295}. The gas density
was derived from the best-fit value of a $\beta$-model to the
surface density profile and the temperature was assumed to be 7 keV.
$y$ turns out to be 1.3 $~10^{-4}$. For S1077 and S295 the X-ray
properties listed in table 1 were taken from the ROSAT All Sky
Survey and the luminosities were estimated assuming $T_e$ = 7 keV.

We have checked that the selected candidates do not contain IR sources of the
IRAS Faint Source Catalogue (\cite{MOS89}).
But sources
fainter than the 90 \% completeness limit of this survey
($\sim$ 230 mJy at 60 $\mu m$) may fall in our beam.
On the basis of our observations on galaxies
performed at SEST and IRAM (\cite{AF92} and \cite{AF95}
), the expected millimetric emission, scaled at the cluster redshift,
is less than 0.8 mJy at 1.2 {\it mm}. This flux would
produce an output signal in antenna temperature of 0.03 mK, which is
lower than the expected signals from the S-Z effect (see below).

The magnitude of the expected signals from the effect
was estimated by
convolving ${\Delta T \over T} = ~y~(x ~{{ e^x +1 }\over {e^x -1}} -4) $
with the spectral filter response,
the atmospheric transmission and the beam-shape,
where T is the CMB temperature, $x=h\nu/kT$ and $y=\int (kT_e/mc^2)~
n_e \sigma _T d\ell$ is the comptonization parameter, $n_e$, T$_e$ being
the electron density and temperature.
We find: ${\Delta T \over T} \sim 0.69 y $, ${\Delta T \over T} \sim - 1.77 y$
at 1.2 and 2 {\it mm} respectively for the thermodynamic
temperatures.

In order to reduce the atmospheric noise and systematics from the antenna
two combined observing strategies were exploited.

\noindent
(1) To reduce temporal and spatial drifts of the atmosphere
a three beams
technique (beam switching + nodding) was applied.
Let us call position A that with the source in the right beam and
the reference beam on the left and position B that with the source
in the left beam with the reference beam on the right, the antenna
tracks each position to integrate 10s and moves according to
the following temporal sequence ABBAABBAA....A.
20 positions are tracked so that the entire cycle lasts 200s, which
is the shortest integration time and which in the following
will be called {\it one scan} (see \S 4 for further details). The duty
cycle was 80 \%, i.e. additional 4s are spent
for each movement of the antenna between position A and B. 

\noindent
(2) In order to properly quantify the contribution to the source signals
from unknown systematics we  measured them using the following procedure.
Each source was integrated over time chunks of 600s (3 of the above scans),
this time interval plus
the needed overheads gives a total tracking time on the source of 15 minutes.
The same time was spent on a blank sky located in equatorial coordinates 
15 minutes ahead in right ascension. This means that the antenna tracks
twice the same sky position in horizontal coordinates: once ON the source
and the other OFF 
the source. This enables us to compare the two different measurements and
quantify the spurious signals (atmosphere + antenna systematics).
The choice of 600s of integration ON and OFF the source is a compromise
between the minimization of the time wasted on overheads and the need of
minimizing the atmospheric variations between one observation and the other.

\section{Data Analysis}

As described in \S 3 the data consist of chunks of 3 {\it scans}, each
containing 20 {\it sub-scans} of 10s corresponding to either position A or
B. The reduction procedure is done in several steps.
(1) One scan data (200s: 440 data points) are  roughly Gaussian
distributed with a
width larger than the detector noise because of the
atmospheric noise. In order to get rid of individual data inconsistent with the
underlying Gaussian distribution, each datum with a value exceeding
3 times the r.m.s. of the entire scan was rejected.
This criterion
eliminates spikes due to equipment malfunction or atmospheric sudden
variation.
The fraction of rejected data is 1\%.
(2) Within each {\it sub-scan} high frequency atmospheric noise was
filtered with the Savitzky-Golay
algorithm (Press and Teukolsky, 1990), which smooths the time
series within a moving window
by approximating the underlying function
by a polynomial of $2^{nd} \div 3^{rd}$ order. At each point the
least-squares fit to the n$_L$ + n$_R$ + 1 points in the moving window
is assigned, where n$_L$ and n$_R$ are the parameters chosen
to smooth the data over a bin of 3$t_0$ ($t_0$ being the integration
time of the lock-in amplifiers, i.e. twice the inverse of
the chopping frequency, 4.4Hz). A mean value of the differential
sky temperature ($T^a_A$ or $T^a_B$) is
then obtained for each {\it sub-scan}, whose variance is found with a
procedure of bootstrap resampling (e.g., Barrow et al. 1984).
This method is widely used to take into account correlation among data (induced in this case
by the filtering).
(3) The signal is  obtained by subtracting each couple of subscans
 $\Delta T^a_i = {{T^a_A - T^a_B }
\over {2}}$ with variance $2 \sigma ^2 _i = \sigma^2_A + \sigma^2 _B$.
(4) For each 200s scan the average value is
$ \Delta T^a _m = {{ \sum _i (\Delta T^a_i)/\sigma^2 _i}
\over {\sum _i \sigma^{-2} _i}} $, with a scatter of
$ \sigma^2 _m = { { \sum _i (\Delta T^a_i - \Delta T^a _m)^2 \sigma^{-2} _i}
\over {(\sum _i \sigma^{-2} _i) (M-1) }} $.

\par\noindent
The weighted mean over the 3 scans is
(at the sky position P) $ \Delta T^a _P =
{{ \sum _{m=1} ^3 \Delta T^a _m \cdot \sigma _m ^{-2} } \over
{ \sum _{m=1} ^3 \sigma ^{-2} _m }}  $,
with a scatter about the mean of $ \sigma ^2 _P = 
{{ \sum _{m=1} ^3 (\Delta T^a _m - \Delta T^a _P ) ^2 \cdot \sigma _m ^{-2} }
\over { \sum _{m=1} ^3 \sigma _m ^{-2} }}$.

\par\noindent
For each $\Delta T^a_m$ a weight is assigned: $ w_m = (\sigma^2 _m +
\sigma^2 _P)^{-1}$, where $\sigma ^2 _m$ represents the variance due
to short-term fluctuations, while $\sigma ^2 _P$ that due to medium-term
variations. \par\noindent
The final {\it weighted} mean, over the sky position $P$, is given by:

$$ \langle \Delta T^a \rangle _f =
{{ \sum _{m=1} ^3 \Delta T^a _m \cdot w_m } \over
{ \sum _{m=1} ^3 w_m }} $$

\par\noindent
with estimated variance:

$$ \sigma ^2 _f = {1\over 2}
{{ \sum _{m=1} ^3 (\Delta T^a_m - \langle \Delta T^a \rangle _f) ^2 \cdot w_m }
\over { \sum _{m=1} ^3 w_m }}$$

\noindent
Values obtained when the antenna tracked the source for 600s (3 scans),
hereafter called signal ON,
are subtracted from those taken when the antenna was 15 minutes
ahead in R.A. (hereafter called signal OFF):

$$ \Delta T^a_{SZ} = (\Delta T^a_f)_{ON} - (\Delta T^a _f)_{OFF} $$
\noindent
and the quadratic sum of the two standard deviation has been
used as errorbar:
$ \sigma _{SZ} ^2 = (\sigma ^2 _f )_{ON} + (\sigma ^2 _f )_{OFF} $.
$(\Delta T^a _f)_{ON}$ and $(\Delta T^a_f)_{OFF}$ with their errorbars
are plotted in figure 1 for the three clusters and for
each channel. 

\noindent
The final values, the weighted means over all the subtracted scans in
thermodynamic values and the corresponding $y$ parameters
are listed in table 1.

\section{Discussion}

The size of the $y$ value of A2744 derived in table 1 disagrees with
the estimated one (see \S 3) and cannot be reliably ascribed to
the sole S-Z thermal effect unless some relevant information is missed in the
X-ray data.
The resulting data towards A2744 at 1.2 and 2 {\it mm} and S1077 at
1.2 {\it mm} could be therefore originated from other sources.
Here possible contaminations are briefly discussed.

\subsection{Contamination of kinematic effect}

Peculiar velocities of clusters could
contribute significantly to the signals for values larger than $v_r >
$ 1000 km/s.
 In fact the ratio between the S-Z
kinematic effect, $ {\Delta T \over T} = {v_r\over c} \cdot \tau $
(where $\tau = \int \sigma _T n dl $ is the optical depth for Thomson
scattering along the line of sight and $v_r$ is the peculiar
velocity), and the thermal effect depends on the cluster peculiar
velocity and the gas temperature.
The lack of any other observations of peculiar velocities
on these clusters makes the quantification of this contribution
impossible and on the basis of the present knowledge 
we cannot exclude that a fraction of the signals is due to it,
 even if large peculiar velocities
 seem quite unlikely in most
cosmological scenarios. The
present data can be used to constrain the combination
$\tau \cdot {v_r\over c}$ which turns out to be
$< 3 ~ 10^{-4} $ for A2744, S1077 and S295 (3$\sigma$ values).

\subsection{Contamination of cluster sources}

As mentioned in \S 3 a strong contamination from sources in the
cluster seems to be unlikely: (a) clusters do
not contain many spirals, (b) late-type galaxies at this redshift
would have a millimetric emission below than the expected signals from the
S-Z effect and (c) we are looking at the very center of the cluster
which, as shown from optical images, is devoid of sources. Only the cluster
S295 has three ellipticals very close to the center of the X-ray
image and part of their emission can therefore
falls in our beam. If there was a significant non-thermal emission
we would expect to find positive signals in both channels, contrary to
our results (see table 1).

\noindent
Let us suppose that one or more sources fall in one of the reference beams.
In this case if a mm-emitting source was located in the right beam, we would
find a positive signals in both channels. If it was in the left
reference beam the signals would be negative in both channels.
This seems to be unlikely since we do not find these systematics
(see table 1). However contribution from cluster sources deserves
further investigations and will be the goal of future projects.

\noindent
It is also unlikely that both signals are contaminated
from the emission of unresolved sources: the estimate of
their confusion limit in the beam made
by Franceschini et al. (1991) gives at 1.2 {\it mm}
$\Delta I \sim$ 0.4 mJy, i.e. $\Delta T$ $\sim$ 0.06 mK.

\subsection{Contamination by other spurious signals}

We have considered other effects affecting the signals:
(a) the selected clusters have sizes comparable to our beams and
beam dilution cannot significantly lower the detection probability
(\cite{RAP}),
 but fluxes are reduced up to 30$\div$40 \% because of the limited chop
throw. (b) Diffraction patterns could enter the beams and alter the signals
but constant systematics of
this type should be removed by our observing strategy (see \S 3).
However,
if the antenna did not track precisely the same paths relative to the
local environment, because of a loss of synchronization, some spurious
signals survive and we cannot exclude at present that this never
happened.
(c) We checked if sources are located in the
OFF position (15min ahead in R.A.): from the IRAS Faint Source
Catalogue and 5GHz NRAO survey it turns out that none fall in
these areas. The IRAS 100 $\mu m$ sky maps do not show
any emission from galaxy cirrus at a level greater than
their zero level.

\vskip 1cm
\acknowledgements
The authors are indebted to Glenn Persson 
and to the ESO Workshop team at La Silla.
This work has been partially supported by the P.N.R.A. (Programma
Nazionale di Ricerche in Antartide) and made use of the {\it Skyview} Database,
 developed under NASA ADP Grant NAS5-32068.
Very useful comments by an unknown referee helped to improve this paper.

\clearpage
 
\begin{table*}
\begin{center}
\begin{tabular}{||lll|ll|c||}
\tableline
 & & & & & \\
 \multicolumn{3}{||c} {ROSAT Cluster} & \multicolumn{2}{|c|}{($\Delta
T)_{therm}$}
 & \multicolumn{1}{|c||} {Compt. par.}  \\
 & & & & & \\
 name & redshift &  \multicolumn{1}{c|} {L$_X$(0.1$\div$2.4keV)}
 &  \multicolumn{1}{|c} {1 {\it mm}}
 &  \multicolumn{1}{c|} {2 {\it mm}}  
 & \multicolumn{1}{|c||}{}\\
 &  &  \multicolumn{1}{c|} {($10^{45}$ erg/s)}
 &  \multicolumn{1}{|c} {(mK)}
 &  \multicolumn{1}{c|} {(mK)}  
 & \multicolumn{1}{|c||}{($10^{-4}$)}\\
\tableline
 & & & & & \\
 S1077 & 0.312 & (0.66$\pm$0.16) &
 (0.9 $\pm$ 0.3)  & (-0.8 $\pm$ 0.5) & (5.3$\pm$1.8) \\
 A2744 & 0.308 & (2.00$\pm$0.05)  &
 (0.8 $\pm$ 0.3)  & (-1.4 $\pm$ 0.4) &  (3.8$\pm$1.2)  \\
 S295  & 0.299 & (1.30$\pm$0.15)  &
 (0.1 $\pm$ 0.3) & (-2.5 $\pm$ 1.0) & $<$ 5.3 (3$\sigma$) \\
 & & & & & \\
\tableline
\end{tabular}
\end{center}
\end{table*}

\clearpage

\begin{figure}
\caption{Antenna temperature differences for positions OFF (empty squares)
and ON (filled squares) the source. Right panels refer to 2 {\it mm}
channel, while the left ones to the 1 {\it mm}
channel. Note that in most cases the 2 {\it mm}
signals ON the sources lie below the corresponding one OFF the sources.
The statistical significance is however very low.
The corresponding panels for the 1 {\it mm} channel show the opposite trend
meaning that the signals from the sources are higher than those
OFF the sources.
}
\end{figure}

\clearpage

\begin{thebibliography}{}

\bibitem[Andreani \& Franceschini 1992] {AF92}
Andreani P. \& Franceschini A. 1992, {\astap} 260, 89
\bibitem[Barrow et al. 1984]{B84} Barrow J.D., Bhavsar S.P., \&
Sonoda D.H. 1984, {\mnras} 210, 19p
\bibitem[Birkinshaw 1990]{B90} Birkinshaw M.: 1990, "Observations of the Sunyaev-Zeldovich
effect", in {\it The Cosmic Microwave Background: 25 Years Later}
eds N. Mandolesi \& N. Vittorio, pp 77-94, Kluwer Academic Publisher, Dordrecht
(the Netherlands)
\bibitem[Birkinshaw 1991]{B91}
Birkinshaw M.: 1991, in {\it Physical Cosmology}, ed. A.Blanchard
et al. (Gif-sur-Yvette: Editions Fronti\'eres), p. 177 
\bibitem[Dall'Oglio et al. 1992]{DAL92}
Dall'Oglio et al., 
1992, {\it Exp. Astron.} 2/5, 256
\bibitem[Franceschini et al. 1991]{FR91}
Franceschini A. et al. 1991, {\aap} {\it Sup.}  89, 285-310 
\bibitem[Franceschini \& Andreani 1995]{AF95}
Franceschini A. \& Andreani P. 1995,
{\apj} 440, L5 
\bibitem[Grainge et al. 1993]{GR93}
Grainge K. et al.: 1993, {\mnras} 265, L57
\bibitem[Haslam 1974]{HAS74}
Halsam C.G.T. 1974, {\aaps}, 15, p. 333
\bibitem[Herbig et al. 1995]{HLR} Herbig T., Lawrence C.R., Readhead
A.C.S. and Gulkis S. 1995 {\apj} 449, L5
\bibitem[Klein et al. 1991]{KL91}
Klein U., Rephaeli Y., Schlickeiser R., Wielebinski
R.: 1991, {\astap} 244, 43
\bibitem[Kreysa 1990]{KRE90}
Kreysa E.: 1990, in {\it From Ground-Based to Space-Born Sub-mm
Astronomy}, Proceedings of 29th Li\`ege International Astrophysical Colloquium,
 ESA SP-314, p.265-270  
\bibitem[Jones et al. 1993]{J93}
Jones M. et al.: 1993 {\it Nature} 365, 322 
\bibitem[Lemke 1992]{LEM92}
 Lemke R. : 1992, {\it Messung linear polarisierter mm-Strahlung ausegesuchter
Quasars mit dem 30-Meter-IRAM-Teleskop}, Inaugural-Dissertation, Universit\" at
Bonn 
\bibitem[Moshir et al. 1989]{MOS89}
Moshir M. et al., 1989,
{\it Explanatory Supplement to the IRAS Faint Source
Survey}, Pasadena:JPL
\bibitem[Pizzo et al. 1995a]{PIZ95a}  Pizzo L., Andreani P., Dall'Oglio G., Lemke R., Ot\`arola A.,
Whyborn N., 1995a, {\it Exp. Astron.} 6, 249
\bibitem[Pizzo et al. 1995b]{PIZ95b} Pizzo L., Martinis L. \&
Dall'Oglio G.: 1995b, preprint
\bibitem[Press \& Teukolsky 1990]{PT90}
 Press W.H. \& Teukolsky S.A., 1990, {\it Computer in Physics} 4, 669
\bibitem[Rephaeli 1987]{RAP} Rephaeli Y., 1987, {\mnras} 228, 29
\bibitem[Sievers et al. 1991]{SIE91}
 Sievers A.W., Mezger P.G., Gordon M.A., Kreysa E., Haslam C.G.T., Lemke R.:
1991, {\astap}, 251, 231 
\bibitem[Sunyaev \& Zeldovich 1972]{SZ72} Sunyaev R.A. \& Zeldovich
Ya.B. 1972, {\it Comm.Astr.Spa.Phys.} 4, 173 
\bibitem[Sunyaev \& Zeldovich 1981]{SZ81}
Sunyaev R.A. \& Zeldovich Ya.B. 1981, {\it Astrop.Spa.Sci.Rev.} 1, 11 
\bibitem[Ulich 1981]{UL81}  Ulich B.L.: 1981, {\aj}, 86, 1619 
\bibitem[Ulich et al. 1984]{UL84} 
Ulich B.L. et al.: 1984, {\it Icarus} 60, 590 
\bibitem[Wilbanks et al. 1994]{WIL94}
  Wilbanks T.M. et al.: 1994, {\apj}  427, L75 
\bibitem[Zeldovich \& Sunyaev 1969]{SZ69}
  Zeldovich Ya.B. \& Sunyaev R.A., 1969,
{\it Astrop. \& Spa.Sci.} 4, 301
\end{thebibliography}
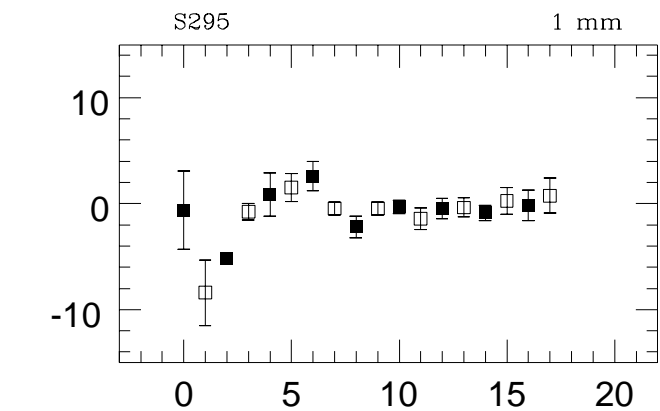
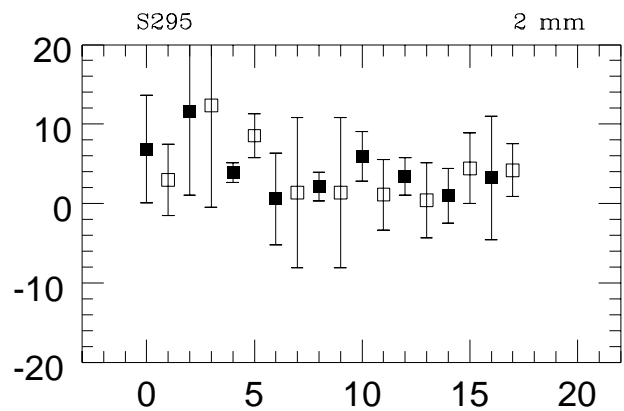
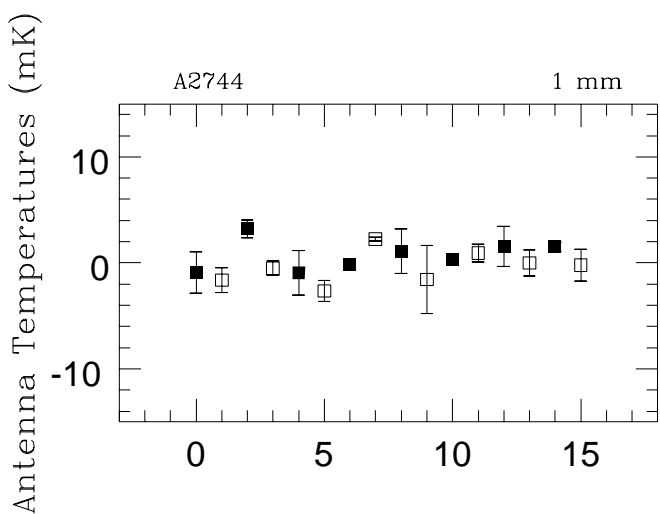
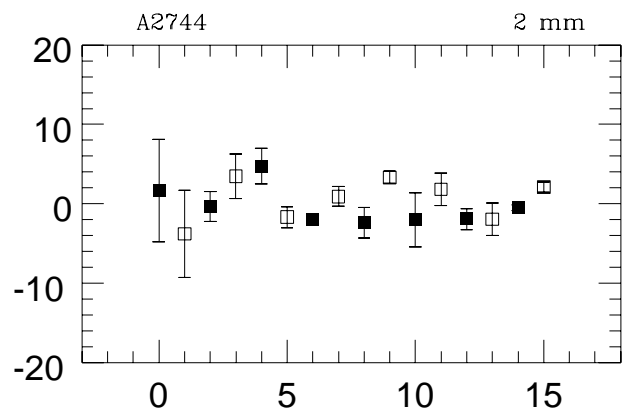
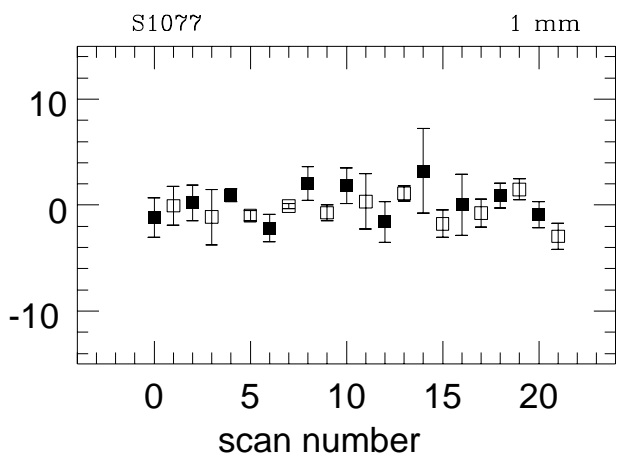
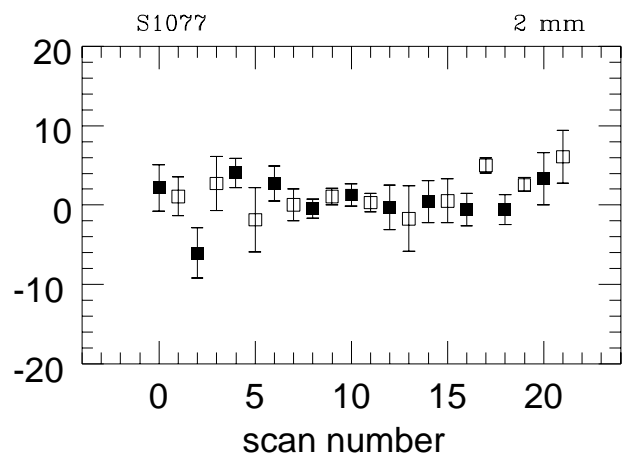
\end{document}